\documentclass[runningheads,a4paper]{llncs}
\usepackage[cp1251]{inputenc}
\usepackage[english]{babel}
\usepackage{amssymb,amsmath}
\usepackage{textcomp}
\usepackage{graphicx}
\usepackage[all]{xy}
\usepackage[usenames]{color}
\usepackage{cite}
\usepackage{dsfont}

\newcommand*\rfrac[2]
{{}^{#1}\!/_{#2}}

\newcommand\bovermat[2]{%
  \makebox[0pt][l]{$\smash{\overbrace{\phantom{%
    \begin{matrix}#2\end{matrix}}}^{\text{#1}}}$}#2}

\newcommand*{\No}{\textnumero}

\begin{document}

\pagestyle{headings}  

\mainmatter              
\title{Protection of Information from Imitation on the Basis of Crypt-Code Structures}
\titlerunning{Protection of Information from Imitation on the Basis of Crypt-Code}  
%
\author{Dmitry Samoylenko\inst{1} \and Mikhail Eremeev\inst{2} \and Oleg Finko\inst{3} \and Sergey Dichenko\inst{3}}
\authorrunning{Dmitry Samoylenko et al.} 
%
\tocauthor{Sergey Dichenko}
\institute{Mozhaiskii Military Space Academy, St. Petersburg, 197198, Russia\\
\email{19sam@mail.ru}\and
Institute a comprehensive safety and special instrumentation of Moscow Technological University, Moscow, 119454, Russia\\
\email{mae1@rambler.ru} \and
Institute of Computer Systems and Information Security of Kuban State Technological University, Krasnodar, 350072, Russia\\
\email{ofinko@yandex.ru}
}
\maketitle
\begin{abstract}
A system is offered for imitation resistant transmitting of encrypted information in wireless communication networks on the basis of redundant residue polynomial codes. The particular feature of this solution is complexing of methods for cryptographic protection of information and multi-character codes that correct errors, and the resulting structures (crypt-code structures) ensure stable functioning of the information protection system in the conditions simulating the activity of the adversary. Such approach also makes it possible to create multi-dimensional ‘‘crypt-code structures’’ to conduct multi-level monitoring and veracious restoration of distorted encrypted information. The use of authentication codes as a means of one of the levels to detect erroneous blocks in the ciphertext in combination with the redundant residue polynomial codes of deductions makes it possible to decrease the introduced redundancy and find distorted blocks of the ciphertext to restore them.

\end{abstract}
\section{Introduction}
The drawback of many modern ciphers used in wireless communication networks is the unresolved problem of complex balanced support of traditional requirements: cryptographic security, imitation resistance and noise stability. It is paradoxical that the existing ciphers have to be resistant to random interference, including the effect of errors multiplication  \cite{Ferg1,Men2,Bur3}. However, such regimes of encrypting as cipher feedback mode are not only the exception, but, on the contrary, initiate the process of error multiplication. The existing means to withstand imitated actions of the intruder, which are based on forming authentication codes and the hash-code~-- only perform the indicator function to determine conformity between the transmitted and the received information \cite{Ferg1,Men2,Chr4}, and does not allow restoring the distorted data.

In some works \cite{MC5,Nid6,SAm7,SAm8} an attempt was made to create the so-called ‘‘noise stability ciphers’’. However, these works only propose partial solutions to the problem (solving only particular types of errors ‘‘insertion’’, ‘‘falling out’’ or ‘‘erasing’’ symbols of the ciphertext etc.), or insufficient knowledge of these ciphers, which does not allow their practical use.
\section{Imitation Resistant Transmitting of Encrypted Information on the Basis of Crypt-Code Structures}
The current strict functional distinction only expects the ciphers to solve the tasks to ensure the required cryptographic security and imitation resistance, while methods of interference resistant coding is expected to ensure noise stability. Such distinction between the essentially inter-related methods to process information to solve inter-related tasks will decrease the usability of the system to function in the conditions of destructive actions of the adversary, the purpose of which is to try to impose on the receiver any (different from the transmitted) message (imposition at random).
At the same time, if these methods are combined, we can obtain both new information ‘‘structures’’~-- crypt-code structures, and a new capability of the system for protected processing of information~-- \textit{imitation resistance} \cite{Petl8}, which we consider to be the ability of the system for \textit{restoration} of veracious encrypted data in the conditions of simulated actions of the intruder, as well as unintentional interference.

The synthesis of crypt-code structures is based on the procedure of complexing of block cypher systems and multi-character correcting codes~\cite{Fin8,Fin9,SAM8}. In one of the variants to implement crypt-code structures as a multi-character correcting code, \textit{redundant residue polynomial codes} (RRPC) can be used, whose mathematical means is based on fundamental provisions of the Chinese remainder theorem for polynomials (CRT)~\cite{Boss9,Mandel10,YuJ-H11}.
\subsection{Chinese Remainder Theorem for Polynomials and Redundant Residue Polynomial Codes}
Let $F[z]$ be ring of polynomials over some finite field $\bbbf_q,$ $q=p^s.$
For some integer $k>1,$ let
$m_1(z),\, m_2(z),\, \ldots,\, m_k(z)\in~F[z]$ be relatively prime polynomials sorted by the increasing degrees,
i.e. $\deg m_1(z) \leq \deg m_2(z) \leq \ldots \leq \deg m_k(z)$, where $\deg m_i(z)$ is the degree of the polynomial.
Let us assume that  $P(z) =\prod_{i=1}^{k}m_i(z).$
Then the presentation of $\varphi$ will establish mutually univocal conformity between polynomials $a(z),$ that do not have a higher degree than
$P(z)$  $\bigl(\deg a(z) < \deg P(z)\bigr)$, and the sets of residues according to the above-described system of bases of polynomials (modules):
\begin{multline*}
\varphi : \rfrac{F[z]}{(P(z))} \rightarrow \rfrac{F[z]}{(m_1(z))} \times \ldots \times \rfrac{F[z]}{(m_k(z))}:\\
: a(z) \mapsto \varphi\bigl(a(z)\bigr) := \bigl(\varphi_1\bigl(a(z)\bigr), \varphi_2\bigl(a(z)\bigr), \ldots, \varphi_k\bigl(a(z)\bigr)\bigr),
\end{multline*}
where $\varphi_i\bigl(a(z)\bigr) := a(z)\mod m_i(z)$ $(i = 1,\, 2,\, \ldots,\, k)$.

In accordance with the CRT, there is a reverse transformation  $\varphi^{-1}$, that makes it possible to transfer the set of residues by the system of bases of polynomials to the positional representation:
\begin{multline}\label{1}
\varphi^{-1} : \rfrac{F[z]}{(m_1(z))} \times \ldots \times \rfrac{F[z]}{(m_k(z))} \rightarrow \rfrac{F[z]}{(P(z))}:\\
: \bigl(c_1(z), \ldots, c_k(z)\bigr) \mapsto a(z) = \sum_{i=1}^{k}c_{i}(z)B_{i}(z)~~{\rm modd}~\bigl(p,~P(z)\bigr),
\end{multline}
where~~ $B_i(z) = k_i(z)P_i(z)$ ~~are~ ~polynomial~ ~~orthogonal ~~bases, ~~$k_i(z) = P^{-1}_i(z)\mod m_i(z),$ $P_i(z) =m_1(z)m_2(z) \ldots m_{i-1}(z)m_{i+1}(z) \ldots m_k(z)$\ \ $(i = 1,\, 2,\, \ldots,\, k).$

Let us also introduce, in addition to the existing number $k,$ the number  $r$ of redundant bases of polynomials while observing the condition of sortednes:
\begin{align}\label{2}
\deg m_1(z)\leq \ldots \leq \deg m_k(z) \leq \deg m_{k+1}(z) \leq \ldots \leq \deg m_{k+r}(z),
\end{align}
and
\begin{align}\label{2.1}
\gcd\bigl(m_i(z),\, m_j(z)\bigr) = 1,
\end{align}
for $i \neq j;$ $ i, j = 1,\, 2,\, \ldots,\, k+r$, then we obtain the expanded RRPC~--- an array of the kind:
\begin{align}\label{3}
C := \bigl(c_1(z),\, \ldots,\, c_k(z),\, c_{k+1}(z),\, \ldots,\, c_n(z)\bigr) : c_i(z) \equiv a(z)\mod m_i(z),
\end{align}
where $n = k + r$,\ $c_i(z) \equiv a(z)\mod m_i(z)$ $(i = 1,\, 2,\, \ldots,\, n)$,\ $a(z) \in \rfrac{F[z]}{(P(z))}$.

Elements of the code $c_i(z)$ will be called symbols, each of which is the essence of polynomials from the quotient ring of polynomials over the module $m_i(z) \in \rfrac{F[z]}{(m_i(z))}.$
At the same time, if
$a(z) \not\in \rfrac{F[z]}{(P(z))},$
then it is considered that this combination contains an error. Therefore, the location of the polynomial $a(z)$ makes it possible to establish if the code combination
$a(z) = \bigl(c_1(z),\, \ldots,\, c_k(z),\, c_{k+1}(z),\, \ldots,\, c_n(z)\bigr)$ is allowed or it contains erroneous symbols.
\subsection{Crypt-Code Structures on Based RRPC}
 Now, the sender-generated message $M$ shall be encrypted and split into blocks of the fixed length $M=\{M_1\|M_2\|\ldots\|M_k\},$ where ‘‘$\|$’’ is the operation of concatenation. Introducing a formal variable $z$ number $i$ block of the open text $M_i,$ we will represent in the polynomial form:
\begin{align*}M_i(z) = \sum_{j=0}^{s-1}m_j^{(i)}z^j = m_{s-1}^{(i)}z^{s-1} + \ldots+m_{1}^{(i)}z+ m_0^{(i)},\end{align*}
where $m_j^{(i)} \in \{0,\, 1\}\ \ \ (i=1,\, 2,\, \ldots,\, k;\ \ \ j = s-1,\, s-2,\, \ldots,\, 0).$

   In order to obtain the sequence of blocks of the ciphertext $\Omega_1(z),\, \Omega_2(z),\, \ldots\\\ldots,\, \Omega_k(z)$ we need to execute $k$
   number of encrypting operations, and to obtain blocks of the open text
    $M_1(z),\, M_2(z),\, \ldots,\, M_k(z),$ we need to execute $k$ number of decrypting operations. The procedures of encrypting and decrypting correspond to the following presentations:
\begin{equation*}
\begin{cases}
\Omega_1(z)\rightarrow{ E_{\kappa_{{\rm e},\,1}}}~:~M_1(z),\\
\Omega_2(z)\rightarrow{ E_{\kappa_{{\rm e},\,2}}}~:~M_2(z),\\
\ldots\ \ldots\ \ldots\ \ldots\ \ldots\ \ldots\  \\
\Omega_k(z)\rightarrow{ E_{\kappa_{{\rm e},\,k}}}~:~M_k(z);\\
\end{cases}~~~
\begin{cases}
M_1(z)\rightarrow{ D_{\kappa_{{\rm d},\,1}}}~:~M_1(z),\\
M_2(z)\rightarrow{ D_{\kappa_{{\rm d},\,2}}}~:~M_2(z),\\
\ldots\ \ldots\ \ldots\ \ldots\ \ldots\ \ldots\  \\
M_k(z)\rightarrow{ D_{\kappa_{{\rm d},\,k}}}~:~M_k(z),\\
\end{cases}
\end{equation*}
where $\kappa_{{\rm e},\,i}, \kappa_{{\rm d},\,i}$ are keys (general case) for encrypting and decrypting $(i=1,\, 2,\, \ldots,\, k)$; if $\kappa_{{\rm e},\,i}= \kappa_{{\rm d},\,i}$~--- the cryptosystem is symmetric, if  $\kappa_{{\rm e},\,i} \neq\kappa_{{\rm d},i}$~--- it is asymmetric.

We will express the adopted blocks of the ciphertext and blocks of the open text correspondingly as $\Omega_i^{*}(z)$ and $M_i^{*}(z)$ $(i=1,\, 2,\, \ldots,\, k)$, as they can contain distortions. The formed blocks of the ciphertext $\Omega_i(z)$ will be represented as the minimum residues (deductions) on the pairwise relatively prime polynomials (bases) $m_i(z).$ Here, $\deg \Omega_i(z)<\deg m_i(z).$
The set of blocks of the ciphertext $\Omega_1(z),\, \Omega_2(z),\, \ldots,\, \Omega_k(z)$  will be represented as a single super-block of elements of the RRPC by the system of bases-polynomials $m_1(z), m_2(z), \ldots, m_k(z).$ In accordance with CRT for the set array of polynomials
 $m_1(z),\, m_2(z),\, \ldots,\, m_k(z)$, that meet the condition that $\gcd\bigl(m_i(z),\, m_j(z)\bigr) = 1$, and polynomials  $\Omega_1(z),\, \Omega_2(z),\, \ldots,\, \Omega_k(z)$, such that $\deg \Omega_i(z)<\deg m_i(z)$, the system of congruences
   \begin{equation}\label{4}
\begin{cases}
\Omega(z)\equiv \Omega_1(z)\mod m_1(z),\\
\Omega(z)\equiv\Omega_2(z)\mod m_2(z),\\
\ldots\ \ldots\ \ldots\ \ldots\ \ldots\ \ldots\ \ldots\ \\
\Omega(z)\equiv\Omega_k(z)\mod m_k(z)
\end{cases}
\end{equation}
has the only one solution $\Omega(z)$.

  Then, we execute the operation of expansion (Base Expansion) of the RRPC by introducing $r$ of redundant bases-polynomials $m_{k+1}(z),\, m_{k+2}(z),\, \ldots\\\ldots, \, m_{k+r}(z)$ that meet the condition (\ref{2}), (\ref{2.1}) and obtaining in accordance with Eq.~(\ref{3}) redundant blocks of data (residues), which we will express as $\omega_{k+1}(z),\, \omega_{k+2}(z),\, \ldots,\, \omega_{n}(z)$  $(n=k+r)$. The combination of ‘‘informational’’ blocks of the ciphertext and redundant blocks of data form crypt-code structures identified as a code word of the expanded RRPC:
  $\bigl\{\Omega_1(z),\, \ldots,\, \Omega_k(z),\, \omega_{k+1}(z),\,\ldots,\,\omega_{n}(z)\bigr\}_{\text{RRPC}}$.

Here, we define a single error of the code word of RRPC as a random distortion of one of the blocks of the ciphertext; correspondingly the
$b$-fold error is defined as a random distortion of $b$ blocks. At the same time, it is known that RRPC detects $b$ errors, if $r\geq b$, and will correct $b$ or less errors, if  $2b\leq r$~\cite{Fin8,Boss9,Mandel10}.

   The adversary, who affects communication channels, intercepts the information or simulates false information. At the same time, in order to impose false, as applied to the system under consideration, the adversary has to intercept a set of information blocks of the ciphertext to detect the redundant blocks of data.

    In order to eliminate the potential possibility that the adversary may impose false information, we need to ensure the ‘‘mathematical’’ gap of the procedure (uninterrupted function) of forming redundant elements of code words of the RRPC.  Moreover, code words of RRPC have to be distributed randomly, i.e. uniform distribution of code words in the set array of the code has to be ensured.  In order to achieve that, the formed sequence of redundant blocks of data  $\omega_{j}(z)$  $(j=k+1,\, k+2,\, \ldots,\, n)$ undergoes the procedure of encrypting:
\begin{equation*}
\begin{cases}
\vartheta_{k+1}(z)\rightarrow{ E_{\kappa_{{\rm e},k+1}}}~:~\omega_{k+1}(z),\\
\vartheta_{k+2}(z)\rightarrow{ E_{\kappa_{{\rm e},k+2}}}~:~\omega_{k+2}(z),\\
\ldots\ \ldots\ \ldots\ \ldots\ \ldots\ \ldots\  \\
\vartheta_{n}(z)\rightarrow{ E_{\kappa_{{\rm e},n}}}~:~\omega_{n}(z),\\
\end{cases}
\end{equation*}
where $\kappa_{{\rm e},\,j}\ \ \ (j=k+1,\, k+2,\, \ldots,\, n)$ are the keys for encrypting.

The process of encrypting of redundant symbols of the code word of the RRPC executes transposition of elements of the vector $\bigl \{\omega_{k+1}(z),\, \omega_{k+2}(z),\, \ldots\\ \ldots,\, \omega_{n}(z)\bigl \}\in \mathcal{A}$   onto the formed elements of the vector of redundant encrypted symbols  $\left\{\vartheta_{k+1}(z),\, \vartheta_{k+2}(z),\, \ldots,\, \vartheta_{n}(z)\right\}\in \mathcal{B}$, where $\mathcal{A}$ is the array of blocks of the ciphertext, $\mathcal{B}$ is a finite array.

The operation of transposition excludes the mutually univocal transformation and prevents the adversary from interfering on the basis of the intercepted informational super-block of the RRPC (the ‘‘informational’’ constituent) $\Omega_{i}(z)$  $(i=1,\, 2,\, \ldots,\, k)$ by forming a verification sequence $\omega_{j}(z)$  $(j=k+1,\, k+2,\, \ldots,\, n)$ for overdriving the protection mechanisms and inserting false information. At the same time, it is obvious that, for the adversary, the set of keys $\kappa_{{\rm e},\,j}$ and functions of encrypting $E_{i}(\bullet)$ of the vector of redundant blocks of data forms a certain array $\mathcal{X}$ of the transformation rules, out of whose many variants, the sender and the addressee will only use a certain one \cite{Chr4,Sim12,Zub2017}.

 We should also note the exclusive character of the operation of encrypting the sequence of redundant blocks of data, due to this, its implementation requires a special class of ciphers that do not alter the lengths of blocks of the ciphertext (endomorphic ones) and not creating distortions (like omissions, replacements or insertions) of symbols, for example, ciphers of permutation.
\section{Imitation Resistant Transmitting of Encrypted Information on the Basis of Multidimensional Crypt-Code Structures}
A particular feature of the above-described system is the necessity to introduce redundant encrypted information in accordance with the RRPC characteristics and specified requirements to the repetition factor of the detected or corrected distortions in the sent data. The theory of coding tells us of solutions to obtain quite long interference-resistant codes with good correct ability on the basis of composing shorter codes that allow simpler implementation and are called composite codes \cite{BL13}. Such solutions can be the basis for the procedure to create multidimensional crypt-code structures.

 Similarly to the previous solution, the open text $M$ undergoes the procedure of encrypting. The formed sequence of blocks of the ciphertext $\Omega_1(z),\, \Omega_2(z),\, \ldots,\, \Omega_k(z)$ is split into $k_2$  number of sub-blocks, contain $k_1$  number of blocks of the ciphertext  $\Omega_i(z)$ in each one and it is expressed in the form of a matrix $\mathbf{W}$ sized  $k_1\times k_2$:
\vspace{-4pt}\begin{align*}\label{44}
\mathbf{W}=\left[
  \begin{array}{ccccc}
    \Omega_{1,\, 1}(z) & \Omega_{1,\, 2}(z) &  \ldots & \Omega_{1,\, k_2}(z) \\
    \Omega_{2,\, 1}(z) & \Omega_{2,\, 2}(z) &  \ldots & \Omega_{2,\, k_2}(z) \\
        \vdots &  \vdots &    \ddots &  \vdots \\[-0.2em]
    \Omega_{k_1,\, 1}(z) & \Omega_{k_1,\, 2}(z) &  \ldots & \Omega_{k_1,\, k_2}(z) \\
  \end{array}
\right],
\end{align*}
where the columns of the matrix $\mathbf{W}$ are sub-blocks made of $k_1$ number of blocks of the ciphertext $\Omega_i(z).$

For each line of the matrix $\mathbf{W,}$ redundant blocks of data are formed, for example, using non-binary codes of Reed-Solomon (code RS [particular case]) over  $\bbbf_q,$
that allow the \textit{2-nd} level of monitoring.

The mathematical means of the RS~codes is explained in detail in~\cite{FJ14}, where one of the ways to form it is based on the deriving polynomial
$g(z)$.
 In $\bbbf_q$  the minimal polynomial for any element $\alpha^i$  is equal to  $M^{(i)}=z-\alpha^i$,  then, the polynomial $g(z)$ of the RS~code corresponds to the equation:
\begin{equation}\label{44}
g(z)=\bigl(z-\alpha^{t}\bigr)\bigl(z-\alpha^{t}\bigr)\ldots\bigl(z-\alpha^{t+2b-1}\bigr),
\end{equation}	 	
where $2b=n-k$; usually $t=0$ or $t=1$.

At the same time, the RS~code is cyclic and the procedure of forming the systematic RS~code is described by the equation:
\begin{align}\label{45}
C(z)=U(z)z^{n-k}+R(z),
\end{align}	 	
where $U(z) = u_{k-1}z^{k-1} + \ldots + u_{1}z+u_0$ informational polynomial, and $\{u_{k-1},\, \ldots,\, u_{1},\, u_0\}$ informational code blocks;
$R(z)=h_{r-1}z^{r-1} + \ldots +h_{1}z + h_0$ the residue from dividing the polynomial $U(z)z^{n-k}$ by $g(z)$, a $\{h_{r-1},\,\ldots,\, h_{1},\, h_0\}$ the coefficients of the residue.
 Then the polynomial $C(z)=c_{n-1}z^{n-1} + \ldots +c_{1}z + c_0$
 and, therefore  $\{c_{n-1},\, \ldots,\, c_{1},\, c_0\}= \{u_{k-1},\, \ldots,\, u_{1},\, u_0,\, h_{r-1},\, \ldots,\, h_{1},\, h_0\}$ a code word.

Basing on the primitive irreducible polynomial, setting the characteristic of the field  $\bbbf_q$
 in accordance with the Eq.~(\ref{44}) a deriving polynomial $g(z)$ of the RS code is formed.

Blocks of the ciphertext $\Omega_{i,\,1}(z),$ $\Omega_{i,\,2}(z), \ldots,$ $\Omega_{i,\,k_2}(z)$ are elements $\mathbf{W}$ expressed as elements of the sorted array, at the same time a formal variable $x$ is introduced and a set of ‘‘informational’’ polynomials is formed:
 \begin{align*}
\mho_i(x) = \sum_{j=1}^{k_2}\bigl(\Omega_{i,\,j}(z)\bigr)x^{j-1} = \bigl(\Omega_{i,\,k_2}(z)\bigr)x^{k_2-1}+ \ldots + \bigl(\Omega_{i,\,2}(z)\bigr)x + \Omega_{i,\,1}(z),\end{align*}
where $i=1,\, 2, \ldots,\, k_1$.

For $\mho_i(x)$ $(i=1,\, 2,\, \ldots,\, k_1)$ in accordance with the Eq.~(\ref{45}) a sequence of residues is formed
  \begin{align*}
R_i(x) = \sum_{j=1}^{r_2}\bigl(\omega_{i,\,j}(z)\bigr)x^{j-1} =\bigl(\omega_{i,\,r_2}(z)\bigr)x^{r_2-1} + \ldots +\bigl(\omega_{i,\, 2}(z)\bigr)x + \omega_{i,\,1}(z),\end{align*}
where $\omega_{i,\,j}(z)$~are coefficients of the polynomial $R_i(x)$ $(i=1,\, 2,\, \ldots,\, k_1)$
assumed as redundant blocks of data of the \textit{2-nd} level of monitoring;
$n_2$ is the length of the RS~code, $k_2$ is the number of ‘‘informational’’ symbols (blocks) of the RS~code, $r_2$ is the number of redundant symbols (blocks) of the RS~code; $n_2=k_2+r_2$.

   Matrix $\mathbf{W}$ with generated redundant blocks of data of the \textit{2-nd} level of monitoring will take the form:
 \begin{multline*}
  \mathbf{\Psi} = \mspace{-3mu}\begin{bmatrix}\mathbf{W}_{k_1\times k_2} | \mathbf{\Upsilon}_{k_1\times r_2}\end{bmatrix}=\mspace{-3mu}
   \begin{matrix}
  \begin{bmatrix}
  ~\bovermat{$~~~~k_2~~~~$}
 {{ \Omega_{1,\,1}(z)} \mspace{-3mu}& { \ldots}& \mspace{-2mu} {\Omega_{1,\,k_2}(z)}} \mspace{-2mu}& \bovermat{$r_2$}{{ \omega_{1,\,k_2+1}(z)} \mspace{-2mu}& { \ldots} & \mspace{-3mu}{\omega_{1,\,n_2}(z)}} \\
 { \Omega_{2,\,1}(z)}  \mspace{-3mu}&{ \ldots}&\mspace{-2mu} {\Omega_{2,\,k_2}(z)}   \mspace{-2mu} & { \omega_{2,\,k_2+1}(z)} \mspace{-2mu}& {\ldots} &\mspace{-3mu} { \omega_{2,\,n_2}(z)} \\
 { \cdots }          \mspace{-3mu}&{  \cdots}& \mspace{-2mu} { \cdots  }            & { \cdots}             \mspace{-2mu}& { \cdots} & \mspace{-3mu}{  \cdots  }               \\[0.1em]
 { \Omega_{k_1,\,1}(z)} \mspace{-3mu}&{ \ldots}& \mspace{-2mu}{\Omega_{k_1,\,k_2}(z)}  \mspace{-2mu}  & { \omega_{k_1,\,k_2+1}(z)}\mspace{-2mu} & {\ldots} &\mspace{-3mu} { \omega_{k_1,\,n_2}(z)} \\
  \end{bmatrix}
  \begin{aligned}
  &\mspace{-13mu}\left.\begin{matrix}
    \\[0.1em]
    \\[0.10em]
    \\[0.10em]
    \\[0.10em]
  \end{matrix} \right\} %
  {\scriptstyle k_1}\\
   \end{aligned}.
 \end{matrix}
 \end{multline*}

 The lines of the matrix  $\mathbf{\Upsilon}$ are redundant blocks of data of the \textit{2-nd} level of monitoring that undergo the procedure of encrypting:
 \begin{equation*}
\begin{cases}
\vartheta_{1,\,\gamma}(z)\rightarrow E_{\kappa_{{\rm e}_{1,\,\gamma}}}~:~ \omega_{1,\,\gamma}(z),\\[-0.2em]
\vartheta_{2,\,\gamma}(z)\rightarrow E_{\kappa_{{\rm e}_{2,\,\gamma}}}~:~ \omega_{2,\,\gamma}(z),\\[-0.2em]
\ldots\ \ldots\ \ldots\ \ldots\ \ldots\ \ldots\  \ldots\ \\[-0.2em]
\vartheta_{k_1,\,\gamma}(z)\rightarrow E_{\kappa_{{\rm e}_{k_1,\,\gamma}}}~:~ \omega_{k_1,\,\gamma}(z),\\[-0.2em]
\end{cases}
\end{equation*}
where $\kappa_{{\rm e}_{i,\,\gamma}}~(i=1,\, 2,\, \ldots,\, k_1;\ \ \ \gamma=k_2+1,\, k_2+2,\, \ldots,\, n_2)$ are the keys for encrypting.

The generated sequence of blocks of the redundant ciphertext of the \textit{2-nd} level of monitoring $\vartheta_{i,k_2+1}(z),$ $\vartheta_{i,k_2+2}(z),$ $\ldots,$ $\vartheta_{i,n_2}(z)$ $(i=1, 2, \ldots, k_1)$ form a matrix $\mathbf{V}$ sized $k_1\times r_2$ redundant blocks of the ciphertext of the \textit{2-nd} level of monitoring:
   \begin{align*}
\mathbf{V}=\left[
  \begin{array}{ccccc}
   \vartheta_{1,\, k_2+1}(z) & \vartheta_{1,\, k_2+2}(z) & \ldots & \vartheta_{1,\, n_2}(z) \\[-0.1em]
    \vartheta_{2,\, k_2+1}(z) & \vartheta_{2,\, k_2+2}(z) &  \ldots & \vartheta_{2,\, n_2}(z) \\[-0.1em]
    \ldots& \ldots & \ldots &\ldots \\[-0.1em]
  \vartheta_{k_1,\, k_2+1}(z) &  \vartheta_{k_1,\, k_2+2}(z)&  \ldots & \vartheta_{k_1,\, n_2}(z) \\[-0.1em]
  \end{array}
\right].
\end{align*}

Now, each column of the matrix $\mathbf{W}$ and $\mathbf{V}$ as a sequence of blocks of the ciphertext $\Omega_{1,\,j}(z),\, \Omega_{2,\,j}(z),\, \ldots,\, \Omega_{k_1,\,j}(z)\ \ (j=1,\, 2,\, \ldots,\, k_2)$
 and
   $\vartheta_{1,\,\gamma}(z),\, \vartheta_{2,\,\gamma}(z),\, \ldots,\,\vartheta_{k_1,\,\gamma}(z)\ \  (\gamma=k_2+1,\, k_2+2,\, \ldots,\, n_2)$
  are expressed in the form of minimal residues on the bases-polynomials $m_i(z)$,  such that $\gcd\bigl(m_i(z),\, m_j(z)\bigr)= 1$ $(i\neq j;\ \ \ i,\,j = 1,\, 2,\, \ldots,\, k_1).$ At the same time  $\deg \Omega_{i,\,j}(z) < \deg m_i(z)$, and $\deg \vartheta_{i,\,\gamma}(z) < \deg m_{i}(z)$.
  Then, as we have noted above, the arrays of blocks of the ciphertext  $\Omega_{1,\,j}(z), \Omega_{2,\,j}(z),\,\ldots,\, \Omega_{k_1,\,j}(z)$ $(j=1,\, 2,\, \ldots,\, k_2)$ and
  $\vartheta_{1,\,\gamma}(z),\, \vartheta_{2,\,\gamma}(z),\,\ldots,\, \vartheta_{k_1,\,\gamma}(z)$ $(\gamma=k_2+1,\, k_2+2,\, \ldots,\, n_2)$
     are expressed as united informational super-blocks of RRPC on the system of bases $m_1(z), m_2(z),\ldots, m_{k_1}(z)$. In accordance with CRT for the specified array of polynomials $m_1(z),\, m_2(z),\,\ldots, m_{k_1}(z)$ that meet the condition $\gcd\bigl(m_i(z),\, m_j(z)\bigr)= 1$, polynomials $\Omega_{1,\,j}(z), \Omega_{2,\,j}(z),\,\ldots,\, \Omega_{k_1,\,j}(z)$  $(j=1, 2, \ldots, k_2)$
     and $\vartheta_{1,\gamma}(z), \vartheta_{2,\,\gamma}(z),\ldots, \vartheta_{k_1,\,\gamma}(z)$ $(\gamma=k_2+1, k_2+2, \ldots, n_2)$
          such that
 $\deg \Omega_{i,\,j}(z)<\deg m_i(z)$, $\deg \vartheta_{i,\,\gamma}(z)<\deg m_i(z)$, the system of congruences (\ref{4}) will take the form:
 \vspace{-3pt} \begin{equation}\label{7}
  \begin{cases}
\begin{cases}
\Omega_1(z)\equiv \Omega_{1,\,1}(z)\mod m_1(z), \\[-0.2em]
\Omega_1(z)\equiv \Omega_{2,\,1}(z)\mod m_2(z), \\[-0.2em]
 \ldots\ \ldots\ \ldots\ \ldots\ \ldots\ \ldots\ \ldots\ \ldots \\[-0.2em]
\Omega_{1}(z)\equiv \Omega_{k_1,\,1}(z)\mod m_{k_1}(z);\\[-0.2em]
\end{cases}\\
~~~~\ldots\ \ldots\ \ldots\ \ldots\ \ldots\ \ldots\ \ldots\ \ldots \\[-0.2em]
\begin{cases}
\Omega_{k_2}(z)\equiv \Omega_{1,\,k_2}(z)\mod m_1(z), \\[-0.2em]
\Omega_{k_2}(z)\equiv \Omega_{2,\, k_2}(z)\mod m_2(z), \\[-0.2em]
 \ldots\ \ldots\ \ldots\ \ldots\ \ldots\  \ldots\ \ldots\ \ldots \\[-0.2em]
\Omega_{k_2}(z)\equiv \Omega_{k_1,\,k_2}(z)\mod m_{k_1}(z);\\[-0.2em]
\end{cases}
\end{cases}
\end{equation}\vspace{-5pt}
\begin{equation}\label{8}
  \begin{cases}
\begin{cases}
\vartheta_{k_2+1}(z)\equiv \vartheta_{1,\,k_2+1}(z)\mod m_1(z), \\[-0.2em]
\vartheta_{k_2+1}(z)\equiv \vartheta_{2,k_2+1}(z)\mod m_2(z), \\[-0.2em]
 \ldots\ \ldots\ \ldots\ \ldots\ \ldots\ \ldots\  \ldots\ \ldots\ \ldots \\[-0.2em]
\vartheta_{k_2+1}(z)\equiv \vartheta_{k_1,\,k_2+1}(z)\mod m_{k_1}(z);\\[-0.2em]
\end{cases}\\
~~~~\ldots\ \ldots\ \ldots\ \ldots\ \ldots\ \ldots\ \ldots\  \ldots \\[-0.2em]
\begin{cases}
\vartheta_{n_2}(z)\equiv \vartheta_{1,\,n_2}(z)\mod m_1(z), \\[-0.2em]
\vartheta_{n_2}(z)\equiv \vartheta_{2,\, n_2}(z)\mod m_2(z), \\[-0.2em]
 \ldots\ \ldots\  \ldots\ \ldots\ \ldots\ \ldots\  \ldots\  \ldots \\[-0.2em]
\vartheta_{n_2}(z)\equiv \vartheta_{k_1,\,n_2}(z)\mod m_{k_1}(z),\\[-0.2em]
\end{cases}
\end{cases}
\end{equation}
 where $\Omega_j(z)$,  $\vartheta_\gamma(z)$ are the only solutions for $j = 1, 2, \ldots, k_2;
  \gamma = k_2+1, \ldots, n_2$.

 Now, according to the additionally formed $r_1$ redundant bases of polynomials $m_{k_1+1}(z),\, m_{k_1+2}(z),\, \ldots,\, m_{n_1}(z)$ $(n_1=k_1+r_1)$, meeting the condition (\ref{2}), (\ref{2.1}) and in accordance with the Eq.~(\ref{3}) redundant blocks of data are formed, that belong to the \textit{1-st} level of monitoring, expressed as  $\omega_{k_1+1,\,j}(z)$,  $\omega_{k_1+2,\,j}(z),\, \ldots,\,\omega_{n_1,\,j}(z)$ $(j = 1,\, 2,\, \ldots,\, k_2)$, as well as reference blocks of data
 $\omega_{k_1+1,\,\gamma}(z)$, $\omega_{k_1+2,\,\gamma}(z),\, \ldots,\, \omega_{n_1,\,\gamma}(z)$ $(\gamma = k_2+1,\,  k_2+2\, \ldots,\, n_2)$.

  The formed redundant blocks of data o the \textit{1-st} level of monitoring  $\omega_{k_1+1,\,j}(z)$, $\omega_{k_1+2,\,j}(z),\, \ldots,\, \omega_{n_1,\,j}(z)$ $(j = 1,\, 2,\, \ldots,\, k_2)$
        are encrypted:
        \begin{equation*}
\begin{cases}
\vartheta_{k_1+1,\,\gamma}(z)\rightarrow E_{\kappa_{{\rm e}_{k_1+1,\,\gamma}}}~:~ \omega_{k_1+1,\,\gamma}(z),\\
\vartheta_{k_1+2,\,\gamma}(z)\rightarrow E_{\kappa_{{\rm e}_{k_1+2,\,\gamma}}}~:~ \omega_{k_1+2,\,\gamma}(z),\\
\ldots\ \ldots\ \ldots\ \ldots\ \ldots\ \ldots\  \ldots\ \ldots\  \\
\vartheta_{n_1,\,\gamma}(z)\rightarrow E_{\kappa_{{\rm e}_{n_1,\,\gamma}}}~:~~ \omega_{n_1,\,\gamma}(z),
\end{cases}
\end{equation*}
 where $\kappa_{{\rm e}_{\iota,\,\gamma}}~(\iota=k_1+1,\, k_1+2,\, \ldots,\, n_1; \gamma=k_2+1,\, k_2+2,\, \ldots,\, n_2)$ are the keys for encrypting.

 Now, the arrays of informational blocks of the ciphertext $\Omega_1(z),\,\Omega_2(z),\, \ldots\\\ldots, \,\Omega_k(z)$, blocks of the redundant encrypted text of the \textit{1-st} and \textit{2-nd} levels of monitoring $\vartheta_{k_1+1,\,j}(z)$, $\vartheta_{k_1+2,\,j}(z),\, \ldots,\, \vartheta_{n_1,\,j}(z)$ $(j=1,\, 2,\, \ldots,\, k_2)$  and $\vartheta_{i,\,k_2+1}(z)$, $\vartheta_{i,\,k_2+2}(z),\, \ldots,\,\vartheta_{i,\,n_2}(z)$
 $(i=1,\, 2,\, \ldots,\, k_1)$,  as well as reference blocks of data $\omega_{k_1+1,\,\gamma}(z),\, \omega_{k_1+2,\,\gamma}(z),\, \ldots,\, \omega_{n_1,\,\gamma}(z)$ $(\gamma = k_2+1,  k_2+2\ldots, n_2)$ form multidimensional crypt-code structures, whose matrix representation correspond to the expression:
  \\[0.2em]\begin{equation*}
\mathbf{\Phi}=
 \begin{matrix}
  \begin{bmatrix}
  ~\bovermat{$~~~~~~~~~k_2~~~~~~~~~$}
 {~~~{ \Omega_{1,\,1}(z)}  &{ \ldots}  &  {\Omega_{1,\,k_2}(z)~~~}} & \bovermat{$~~~~r_2~~~~$}{{ \vartheta_{1,\,k_2+1}(z)} & { \ldots} & {\vartheta_{1,\,n_2}(z)~~~~}} \\
%
%
 { \ldots}         &{ \ldots}&   { \ldots}              & { \ldots}              & { \ldots} &  { \ldots}               \\
{ \Omega_{k_1,\,1}(z)}  &{ \ldots}  & {\Omega_{k_1,\,k_2}(z)}    &{\vartheta_{k_1,\,k_2+1}(z)} & { \ldots} &  {\vartheta_{k_1,\,n_2}(z) } \\
 \\
  {\vartheta_{k_1+1,\,1}(z)}    &{ \ldots}&   {\vartheta_{k_1+1,\,k_2}(z)}     & {\omega_{k_1+1,\,k_2+1}(z)} & { \ldots} & { \omega_{k_1+1,\,n_2}(z)} \\
  %
  %
  { \ldots}             &{\ldots}&  {\ldots}             & { \ldots}              & { \ldots} &  { \ldots}               \\
   {\vartheta_{n_1,\,1}(z)}   &{\ldots}& {\vartheta_{n_1,\,k_2}(z)}    & {\omega_{n_1,\,k_2+1}(z)}  & { \ldots} &  {\omega_{n_1,\,n_2}(z)} \\
  \end{bmatrix}
  \begin{aligned}
 \mspace{-1mu} &\left.\begin{matrix}
    \\[0.2em]
    \\[0.2em]
    \\[0.2em]
  \end{matrix} \right\} %
  {\scriptstyle k_1}\\
  \mspace{-1mu}   &\left.\begin{matrix}
    \\[0.2em]
    \\[0.2em]
    \\[0.2em]
  \end{matrix} \right\} %
 {\scriptstyle r_1}\\
   \end{aligned}
 \end{matrix}.
  \end{equation*}

The formed multidimensional crypt-code structures correspond to the following parameters (a particular case for 2 levels of monitoring):
\begin{align*}\label{172}
\left\{
\begin{aligned}
&n=n_1n_2, \\[-0.2em]
&k=k_1k_2, \\[-0.2em]
&r=r_1n_2 + r_2n_1 -r_1r_2,\\[-0.2em]
&d_{\min}=d_{\min_1}d_{\min_2},\\[-0.2em]
\end{aligned}
\right.
\end{align*}
where $n,$ $k,$ $r,$ $d_{\min}$ are generalized monitoring parameters; $n_i,$ $k_i,$ $r_i,$ $d_{\min_i}$ are parameters of the level of monitoring number $i$ $(i=1,\, 2)$ \cite{BL13}.

On the receiving side, multidimensional crypt-code structures undergo the procedure of reverse transformation. In order to achieve that, the received sequence of blocks of the ciphertext $\Omega_{i}(z)$ $(i = 1,\, 2,\, \ldots,\, k)$ is split into $k_2$ number of sub-blocks containing $k_1$ blocks of the ciphertext and expressed in the form of the matrix $\mathbf{W}^*$ with the parameters identical to the parameters of the sending side:
\begin{equation*}\label{448}
\mathbf{W^*}=\left[
  \begin{array}{ccccc}
    \Omega^*_{1,\, 1}(z) & \Omega^*_{1,\, 2}(z) &  \ldots & \Omega^*_{1,\, k_2}(z) \\
    \Omega^*_{2,\, 1}(z) & \Omega^*_{2,\, 2}(z) &  \ldots & \Omega^*_{2,\, k_2}(z) \\
        \vdots &  \vdots &    \ddots &  \vdots \\[-0.1em]
    \Omega^*_{k_1,\, 1}(z) & \Omega^*_{k_1,\, 2}(z) &  \ldots & \Omega^*_{k_1,\, k_2}(z) \\
  \end{array}
\right],
\end{equation*}
where the columns of the matrix $\mathbf{W}^*$ are sub-blocks of $k_1$ blocks of the ciphertext $\Omega_i^*(z)$.
The arrays of blocks of the redundant ciphertext of the \textit{1-st} and \textit{2-nd} levels of monitoring  $\vartheta^*_{k_1+1,\,j}(z),\, \vartheta^*_{k_1+2,\,j}(z),\, \ldots,\,  \vartheta^*_{n_1,\,j}(z)$ $(j=1,\, 2,\, \ldots,\, k_2)$,\, $\vartheta^*_{i,\,k_2+1}(z),\,\vartheta^*_{i,\,k_2+2}(z),\, \ldots,\, \vartheta^*_{i,\,n_2}(z)$ $(i=1,\, 2,\, \ldots,\, k_1)$
  that were obtained in the parallel process undergo procedure of decrypting:
 \begin{equation*}
\begin{cases}
\omega^{*}_{k_1+1,\,j}(z)\rightarrow D_{\kappa_{{\rm d}_{k_1+1,\,j}}}~:~ \vartheta_{k_1+1,\,j}^*(z),\\
\omega^{*}_{k_1+2,\,j}(z)\rightarrow D_{\kappa_{{\rm d}_{k_1+2,\,j}}}~:~ \vartheta_{k_1+2,\,j}^*(z),\\
\ldots\ \ldots\ \ldots\ \ldots\ \ldots\ \ldots\  \ldots\ \ldots\    \\[-0.2em]
\omega^{*}_{n_1,\,j}(z)\rightarrow D_{\kappa_{{\rm d}_{n_1,\,j}}}~:~ \vartheta_{n_1,\,j}^*(z);\\
\end{cases}
\begin{cases}
\omega^{*}_{1,\,\gamma}(z)\rightarrow D_{\kappa_{{\rm d}_{1,\,\gamma}}}~:~ \vartheta_{1,\,\gamma}^*(z),\\
\omega^{*}_{2,\,\gamma}(z)\rightarrow D_{\kappa_{{\rm d}_{2,\,\gamma}}}~:~ \vartheta_{2,\,\gamma}^*(z),\\
\ldots\ \ldots\ \ldots\ \ldots\ \ldots\ \ldots\  \ldots\ \\[-0.2em]
\omega^{*}_{k_1,\,\gamma}(z)\rightarrow D_{\kappa_{{\rm d}_{k_1,\,\gamma}}}~:~ \vartheta_{k_1,\,\gamma}^*(z),\\
\end{cases}
\end{equation*}
where $\kappa_{{\rm d}_{\iota,\,j}}$ and $\kappa_{{\rm d}_{i,\,\gamma}}$
$(\iota=k_1+1,\, k_1+2,\, \ldots,\, n_1;\ \ j=1,\, 2,\, \ldots,\, k_2)$,\\
$(i=1,\, 2,\, \ldots,\, k_1;\ \  \gamma=k_2+1,\, k_2+2,\, \ldots,\, n_2)$  are the keys for decrypting.

 Now, every column $\Omega^*_{1,\,j}(z),\, \Omega^*_{2,\,j}(z),\, \ldots\, \Omega^*_{k_1,\,j}(z)$ of the matrix $\mathbf{W}^*$ that is interpreted as an informational super-block of the RRPC is put into the conformity to the sequence of redundant blocks of data of the \textit{1-st} level of monitoring $\omega^*_{k_1+1,\,j}(z),\, \omega^*_{k_1+2,\,j}(z)\,\ldots,\, \omega^*_{n_1,\,j}(z)$ $(j=1,\, 2,\, \ldots,\, k_2)$ on the bases-polynomials  $m_i(z)\ \   (i=1,\,2,\, \ldots,\, n_1)$ resulting in forming the code vector of the expanded RRPC $\bigl\{\Omega^*_{1,\,j}(z),\, \ldots,\, \Omega^*_{k_1,\,j}(z),\, \omega^*_{k_1+1,\,j}(z),\, \ldots,\, \omega^*_{n_1,\,j}(z)\bigr\}_{\text{RRPC}}$.

Besides that, the columns of the \textit{2-nd} level of monitoring $\vartheta^*_{1,\gamma}(z), \ldots, \vartheta^*_{k_1,\gamma}(z)$  are put into the conformity to the reference blocks of data   $\omega^*_{k_1+1,\gamma}(z), \ldots,  \omega^*_{n_1,\gamma}(z)$  $(\gamma=k_2+1, \ldots, n_2)$ on the bases-polynomials  $m_i(z) (i=1,2, \ldots, n_1)$  and a code vector of the expanded RRPC $\bigl\{\vartheta^*_{1,\gamma}(z), \ldots, \vartheta^*_{k_1,\gamma}(z), \omega^*_{k_1+1,\gamma}(z), \ldots\\
\ldots, \omega^*_{n_1,\gamma}(z)\bigr\}_{\text{RRPC}}$ is formed. Then, the procedure is started to detect the RRPC elements distorted (simulated) by the adversary, basing on the detection capability conditioned by the equation $d_{\min_1}-1$. At the same time, if  $\Omega_j^*(z), \vartheta_\gamma^*(z)\in\rfrac{F[z]}{(P(z))},$  then we assume that there are no distorted blocks of the ciphertext, where $\Omega_j^*(z), \vartheta_\gamma^*(z)$ solution of the comparison system  (\ref{7}), (\ref{8}) in accordance with the Eq.~(\ref{3}), for $j=1, 2, \ldots, k_2; \gamma=k_2+1, \ldots, n_2$. Considering the condition $\left\lfloor(d_{\min_1}-1)2^{-1}\right\rfloor$, the procedure of restoring the distorted elements of RRPC can be executed with the help of calculating the minimal residues
or with any other known method of RRPC decoding.

The corrected (restored) elements number  $j$ of the sequence of the ciphertext blocks $\Omega^{**}_{1,j}(z), \Omega^{**}_{2,j}(z), \ldots, \Omega^{**}_{k_1,j}(z)$ ‘‘replace’’ the distorted number $i$ (of the ciphertext blocks) of the lines $\Omega^{*}_{i,1}(z), \Omega^{*}_{i,2}(z), \ldots, \Omega^{*}_{i,k_2}(z)$   $(i=1,2,\ldots,k_1)$  of the matrix $\mathbf{W}^*.$ The symbols  ‘‘**’’ indicate the stochastic character of restoration.

Now, each line  $\Omega^{*}_{i,1}(z), \Omega^{*}_{i,2}(z), \ldots, \Omega^{*}_{i,k_2}(z)$  is put into conformity of the blocks of the redundant ciphertext of the \textit{2-nd} level of monitoring $\omega^{*}_{i,k_2+1}(z),$ $\omega^{*}_{i,k_2+2}(z), \ldots,$ $\omega^{*}_{i,n_2}(z)$  $(i=1,2,\ldots,k_1)$  and code vectors are formed for the RS~code  $\bigl\{\Omega^*_{i,1}(z), \ldots,$ $\Omega^*_{i,k_2}(z), \omega^*_{i,k_2+1}(z),$ $\ldots,$ $ \omega^*_{i,n_2}(z)\bigr\}_{\text{RS}}$.

According to the code vectors, polynomials are formed
\begin{align*}
\mathcal{C}_i^{*}(x)=\mho_i^{*}(x)+R_i^{*}(x) = \sum_{j=1}^{k_2}\bigl(\Omega_{i,j}^{*}(z)\bigr)x^{j-1}+\sum_{\gamma=k_2+1}^{n_2}\bigl(\omega_{i,\gamma}^{*}(z)\bigr)x^{\gamma-1}
\end{align*}
and their values are calculated for the degrees of the primitive element of the field $\alpha^\ell:$
\begin{align*}
\mathcal{S}_{i,\ell}=\mathcal{C}_i^{*}(\alpha^\ell)= \sum_{j=1}^{k_2}\Bigl(\Omega_{i,j}^{*}(z)\Bigr)\Bigl(\alpha^{(j-1)}\Bigr)^\ell+\sum_{\gamma=k_2+1}^{n_2}\Bigl(\omega_{i,\gamma}^{*}(z)\Bigr)\Bigl(\alpha^{(\gamma-1)}\Bigr)^\ell,
\end{align*}
where  $i=1,\, 2,\, \ldots,\, k_1;~~ \ell=0,\, 1,\, \ldots,\, r_2-1$, $r_2=n_2-k_2$.

At the same time, if the values of checksums $\mathcal{S}_{i,\ell}$ with $\alpha^\ell$ for each vector of the line are equal to zero, then we assume that there are no distortions. Otherwise, the values $\mathcal{S}_{i,\,0}, \mathcal{S}_{i,\,1}, \ldots, \mathcal{S}_{i,\,r_2-1}$ for $i=1,\, 2,\, \ldots,\, k_1$ are used for further restoration of the blocks of the ciphertext
$\Omega_{i,\,1}^{*}(z),$ $\Omega_{i,\,2}^{*}(z), \ldots,$ $\Omega_{i,\,k_2}^{*}(z)$
with the help of well-known algorithms for decoding RS~codes (of Berlekamp-Massey, Euclid, Forney and etc.).

The corrected (restored) sequences of redundant blocks of the ciphertext of the \textit{2-nd} level of monitoring  $\vartheta_{1,\gamma}^{**}(z), \ldots, \vartheta_{k_1,\gamma}^{**}(z)$  are subject of the second transformation (decryption) of redundant blocks of the ciphertext of the \textit{2-nd} ~level of monitoring into redundant blocks of data of the \textit{2-nd} level of monitoring
 $\omega_{1,\gamma}^{**}(z), \ldots, \omega_{k_1,\gamma}^{**}(z)$. The redundant blocks of data of the \textit{2-nd} level of monitoring  $\omega_{1,\gamma}^{**}(z), \ldots, \omega_{k_1,\gamma}^{**}(z)$ $(\gamma=k_2+1, k_2+2, \ldots, n_2)$   that have been formed again are used for forming code combinations of the RS~code and their decoding.

\section{Imitation Resistant Transmitting of Encrypted Information on the Basis of Crypt-Code Structures and Authentication Codes}
Currently, to detect simulation by the adversary in the communication channel, an additional encryption regime is used to simulate imitated insertion (forming an authentication code [Message Authentication Code]) \cite{Ferg1,Men2,Chr4}. A drawback of this method to prevent imitation by the adversary is the lack of possibility to restore veracious information in the systems for transmitting information. Complexing the method to protect from imitating of data on the basis of message authentication codes (MAC) and the above-described solution based on expanding the RRPC with encrypting the redundant information, it shall make it possible to overcome the drawback of the known solution. Let us assume that MAC are formed as usual from the sequence consisting of $k_2$  number of sub-blocks containing $k_1$  blocks each of the ciphertext $\Omega_i(z)$ in each one. Then the procedure of generation of MAC $H_i(z)$   $(i=1, \ldots, k_1)$ can be expressed:
\begin{equation*}\left\{
    \begin{array}{ll}
      H_1(z)  \rightarrow I_{h_{1}}: {\rm {\bf \Omega}}_1, \\[-0.2em]
      H_2(z)  \rightarrow I_{h_{2}}: {\rm {\bf \Omega}}_2, \\[-0.2em]
      \ldots\  \ldots\ \ldots\ \ldots\ \ldots\  \\[-0.2em]
      H_{k_1}(z)  \rightarrow I_{h_{k}}: {\rm {\bf \Omega}}_{k_1},\\[-0.2em]
    \end{array}
  \right.\end{equation*}
where $I_{h_{i}}$  is the operator of generation of an MAC on the key $h_{i}$ $(i=1, \ldots, k_1),$ ${\rm {\bf \Omega}}_i=\bigl\{\Omega_{i, 1}(z), \ldots, \Omega_{i, k_2}(z)\bigr\}$ is a vector equation of the super-block of the ciphertext, $k_2$  is the length of the super-block.
Purposeful interfering of the adversary into the process of transmitting super-blocks of the ciphertext with the MAC calculated from them can cause their distorting.  Correspondingly, on the receiving side, the super-blocks   ${\rm {\bf \Omega}}_i^{*}=\bigl\{\Omega_{i, 1}^{*}(z), \ldots, \Omega_{i, k_2}^{*}(z)\bigr\}$ of the ciphertext are the source for calculating MAC:
\begin{equation*}\left\{
    \begin{array}{ll}
      \widetilde{H}_1(z)  \rightarrow I_{h_{1}}: {\rm {\bf \Omega}}_1^*, \\
      \widetilde{H}_2(z)  \rightarrow I_{h_{2}}: {\rm {\bf \Omega}}_2^*, \\
      \ldots\  \ldots\ \ldots\ \ldots\ \ldots\  \\
      \widetilde{H}_{k_1}(z)  \rightarrow I_{h_{k}}: {\rm {\bf \Omega}}_{k_1}^*,\\
    \end{array}
  \right.\end{equation*}
  where ${\rm {\bf \Omega}}_i^*=\bigl\{\Omega_{i, 1}^*(z), \ldots, \Omega_{i, k_2}^*(z)\bigr\}$ is the received super-block of the ciphertext; $\widetilde{H}_i(z)$ are MAC from the received blocks of the ciphertext, for  $i=1, 2, \ldots, k_1.$

Similarly to the previous solution for restoring the messages simulated by the adversary from the transmitted sequence of blocks of the ciphertext with MAC $\Bigl\{ \bigl\{{\rm {\bf \Omega}}_{1}, H_{1}(z)\bigr\}; \ldots; \bigl\{{\rm {\bf \Omega}}_{k_1}, H_{k_1}(z)\bigr\}; \bigl\{ \vec{\vartheta}_{k_1+1}, H_{k_1+1}(z)\bigr\}; \ldots; \bigl\{ \vec{\vartheta}_{n_1}, H_{n_1}(z)\bigr\}\Bigr\}_{\text{RRPC}}$,
 an extended RRPC is formed.

The sub-system of imitation-resistant reception of encrypted information on the basis of the RRPC and using MAC implements the following algorithm.

\textit{Input:} the received sequence of vectors of encrypted message blocks~with~MAC:
$\Bigl\{ \bigl\{{\rm {\bf \Omega}}_{1}^{*}, H_{1}^{*}(z)\bigr\}; \ldots; \bigl\{{\rm {\bf \Omega}}_{k_1}^{*}, H_{k_1}^{*}(z)\bigl\}; \bigl\{ \vec{\vartheta}_{k_1+1}^{*}, H_{k_1+1}^{*}(z)\bigr\}; \ldots; \bigl\{ \vec{\vartheta}_{n_1}^{*}, H_{n_1}^{*}(z)\bigr\}\Bigr\}_{\text{RRPC}}$.

\textit{Output:} a corrected (restored) array of super-blocks of the ciphertext ${\bf \Omega}^{**}_1, {\bf \Omega}^{**}_2, \ldots, {\bf \Omega}^{**}_{k_1}$.

\textit{Step 1.} Detection of the possible simulation by the adversary in the received sequence of blocks of the ciphertext with localization of the number $i$  row vector
with the detected false blocks of the ciphertext, is executed by comparing the MAC received from the communication channel $H_{1}^{*}(z), \ldots, H_{k_1}^{*}(z),$ $H_{k_1+1}^{*}(z), \ldots,
 H_{n_1}^{*}(z)$ and
 MAC  $\widetilde{H}_{1}^{*}(z), \ldots, \widetilde{H}_{k_1}^{*}(z),$ $\widetilde{H}_{k_1+1}^{*}(z), \ldots,
 \widetilde{H}_{n_1}^{*}(z)$ calculated in the sub-system of data reception.
  Next, a comparison procedure is performed for all row vectors $(i=1,\, \ldots,\, k_1$, $k_1+1,\, \ldots,\, n_1)$:
\begin{equation*}\left\{
  \begin{array}{ll}
  1,~~ \textit{if} ~~\ H_i^*(z)=\widetilde{H}_i(z);\\
  0,~~ \textit{if} ~~\ H_i^*(z)\neq \widetilde{H}_i(z).
      \end{array}
  \right.\end{equation*}

	 	\textit{ Step 2.} Restoring veracious data by solving the congruences systems:
   \begin{equation}\label{11}
\left\{
\begin{array}{l}
\left\{
\begin{array}{l}
\Omega^{**}_1(z)\equiv \Omega^{*}_{{J_1},\,1}(z)\mod m_{{J_1}}(z),\\[-0.1em]
\ldots\   \ldots\ \ldots\ \ldots\ \ldots\ \ldots\ \ldots\ \ldots\  \ldots\  \\[-0.1em]
\Omega^{**}_{1}(z)\equiv \Omega^{*}_{{J_{k_1}},\,1}(z)\mod m_{{J_{k_1}}}(z),\\[-0.1em]
\Omega^{**}_1(z)\equiv \omega^{*}_{{J_{k_1+1}},\,1}(z)\mod m_{{J_{k_1+1}}}(z),\\[-0.1em]
\ldots\   \ldots\ \ldots\ \ldots\ \ldots\ \ldots\ \ldots\ \ldots\  \ldots\  \\[-0.1em]
\Omega^{**}_1(z)\equiv \omega^{*}_{{J_{n_1}},\,1}(z)\mod m_{{J_{n_1}}}(z);\\[-0.1em]
\end{array}
\right. \\[-0.1em]
\left.
\begin{array}{l}
\ldots\   \ldots\ \ldots\ \ldots\ \ldots\ \ldots\ \ldots\ \ldots\  \ldots\   \\
\end{array} \right. \\[-0.2em]
    \left\{
\begin{array}{l}
\Omega^{**}_{k_2}(z)\equiv \Omega^{*}_{{J_1},k_2}(z)\mod  m_{{J_1}}(z),\\
\ldots\   \ldots\ \ldots\ \ldots\ \ldots\ \ldots\ \ldots\ \ldots\  \ldots\  \\[-0.1em]
\Omega^{**}_{k_2}(z)\equiv \Omega^{*}_{{J_{k_1}},k_2}(z)\mod  m_{J_{k_1}}(z),\\
\Omega^{**}_{k_2}(z)\equiv \omega^{*}_{{J_{k_1+1}},k_2}(z)\mod  m_{{J_{k_1+1}}}(z),\\
\ldots\   \ldots\ \ldots\ \ldots\ \ldots\ \ldots\ \ldots\ \ldots\   \ldots\ \\[-0.1em]
\Omega^{**}_{k_2}(z)\equiv \omega^{*}_{{J_{n_1}},k_2}(z)\mod  m_{{J_{n_1}}}(z),\\
\end{array} \right.
\end{array}
\right. \end{equation}
 where $J_1, J_2, \ldots, J_{n_1}$ are row vector numbers, if the comparison result for these MAC showed absence of distortions in
  sequence of blocks of the ciphertext
 ${\rm {\bf \Omega}}^{*}_{j}(z)=\bigl\{\Omega_{j, 1}^*(z), \Omega_{j, 2}^*(z), \ldots,  \Omega _{j, k_2}^*(z)\bigr\}$.
In accordance with the CRT solutions of systems (\ref{11}) is the following:
\begin{multline*}\Omega_{j}^{**}=\Omega_{J_{1}, j}^{*}(z)B_{J_1}(z)+\ldots+\Omega_{J_{k_1}, j}^{*}(z)B_{J_{k_1}}(z)+\ldots\\
\ldots+\omega_{J_{k_1+1}, k}^{*}(z)B_{J_{k_1+1}}(z)+\ldots+\omega_{J_{n_1}, k}^{*}(z)B_{J_{n_1}}(z)~~{\rm modd}~\bigl(p, P_{k_v}(z)\bigr),
  \end{multline*}
where\\
 $\displaystyle B_{J_i}(z)=k_{J_i}(z)P_i(z)$ are polynomial orthogonal bases;
$P_{k_v}(z)=\prod_{\begin{subarray}{l}
i=1, \ldots,k; \\ i\neq v
\end{subarray}}m_i(z)$;
 $v$ is the number of the detected  ‘‘distorted’’ row vector;
$P_{J_i}(z)=P_{k_v}(z)m_i^{-1}(z)$; $ k_{J_i}(z)=P_{J_i}^{-1}(z)\mod m_{J_{i}}(z)$  $(j=1, \ldots, k_2; i=1, \ldots, n_1)$.

The values of polynomial orthogonal bases are calculated beforehand and are stored in the memory of the RRPC decoder. Restoring veracious blocks can be done by calculating the minimal deductions or by any other known method.

In a comparative evaluation of the effectiveness of the methods under consideration for providing imitation resistant transmission of encrypted information, we will assume that the adversary distorts the ciphertext blocks in the generated crypt-code structures with probability $p_{adv}=2\cdot10^{-2}.$ Probability $p_{adv}$ distortion of each ciphertext block is constant and does not depend on the results of receiving the preceding elements of crypt-code structures. The probability $P(b)$ of reception crypt-code structures with $b$ and more errors are presented in the table \ref{tab1}, in accordance with which a higher recovery power is provided multidimensional crypt-code structures (RRP codes and RS codes). At what at the given values $k_1, k_2,$ the closer the matrix being formed  $\mathbf{\Phi}_{n_1 \times n_2}$ to the square shape, the less the level of redundancy introduced.
\begin{table}
\centering
\caption{Effectiveness crypt-code structures  }\label{tab1}
\begin{tabular}{lllllll}
  \hline
  \noalign{\smallskip}
  Method of construction               & ~~Structures~            & ~$n$~  & ~$k$~  & ~$d_{\min}$~  & ~$\frac{k}{n}$~  &  ~$P(b)$ \\
  \noalign{\smallskip}
\hline
\noalign{\smallskip}
  Crypt-code structures                          & ~~(6, 3, 4)~            & ~6~    & ~3~    & ~4~           & ~0.5~           &~0.114158 \\
  \noalign{\smallskip}
\noalign{\smallskip}
  (RRPC)                    & ~~(8, 4, 5)~            & ~8~    & ~4~    & ~5~           & ~0.5~            &~0.010336  \\
  \noalign{\smallskip}
\hline
\noalign{\smallskip}
  Multidimensional crypt-code         & ~~(6, 3, 4); (11, 5, 7)~ & ~66~   & ~15~   & ~28~          & ~0.227~          &~0.000133  \\
  \noalign{\smallskip}
\noalign{\smallskip}
  structures: (RRPC); (RS)                                 & ~~(8, 4, 5); (8, 4, 5)~  & ~64~   & ~16~   & ~25~          & ~0.25~           &~0.000106  \\
  \noalign{\smallskip}
\hline
\noalign{\smallskip}
  Multidimensional crypt-code  & ~~(4, 3, 2); (6, 3, 4)~  & ~24~   & ~9~    & ~8~          & ~0.375~           &~0.008862 \\
  \noalign{\smallskip}
\noalign{\smallskip}
  structures: (MAC); (RRPC)                                     & ~~(4, 3, 2); (8, 4, 5)~  & ~32~   & ~12~    & ~10~          & ~0.375~           &~0.000802  \\
  \hline
\end{tabular}
\end{table}
\section{Conclusion}
The methods of information protection examined in this article (against simulation by the adversary) are based on the composition of block ciphering system and multi-character codes that correct errors by forming crypt-code structures with some redundancy. This redundancy is usually small and it makes it possible to express all the possible states of the protected information. Forming multidimensional crypt-code structures with several levels of monitoring makes it possible to not only detect simulating actions of the intruder but also, if necessary, to restore the distorted encrypted data with the set probability and their preliminary localization.

\end{document}